\documentclass[10pt,aps,prl,reprint,superscriptaddress]{revtex4-1}

\usepackage[utf8]{inputenc}
\usepackage{lipsum}
\usepackage{natbib}
\usepackage{hyperref}
\usepackage{amsmath,hyperref}
\usepackage{amssymb,bm}
\usepackage{enumerate}
\usepackage{float}
\usepackage{color}
\usepackage{graphicx}
\usepackage{color}
\usepackage{dsfont}

\newcommand{\beq}{\begin{equation}}
\newcommand{\eeq}{\end{equation}}
\newcommand{\beqa}{\begin{eqnarray}}
\newcommand{\eeqa}{\end{eqnarray}}
\newcommand{\bea}{\begin{eqnarray}}
\newcommand{\eea}{\end{eqnarray}}

\definecolor{amarillo}{cmyk}{0,0.5,1,0}

\newcommand{\diff}{\text{d}}

\begin{document}  
 
\title{Scale Holography}

\author{José A. R. Cembranos}
\affiliation{Departamento de F\'{\i}sica Te\'orica I, Universidad Complutense 
de Madrid, 28040 Madrid, Spain}
\affiliation{Departamento de  F\'{\i}sica, Universidade de Lisboa, P-1749-016 Lisbon, Portugal}
\author{Salvador E. R. Ciarreta}
\affiliation{Departamento de F\'{\i}sica Te\'orica I, Universidad Complutense 
de Madrid, 28040 Madrid, Spain}
\affiliation{Departamento de F\'{\i}sica Te\'orica II, Universidad Complutense 
de Madrid, 28040 Madrid, Spain}
\author{Luis J. Garay}
\affiliation{Departamento de F\'{\i}sica Te\'orica II, Universidad Complutense 
de Madrid, 28040 Madrid, Spain}
\affiliation{Instituto de Estructura de la Materia (IEM-CSIC), Serrano 121, 28006 Madrid, Spain}

\begin{abstract} 
We present a new correspondence between a $d$-dimensional dynamical system   and a whole family of $(d+1)$-dimensional systems. This new scale-holographic relation is built by the explicit introduction of a dimensionful constant which determines the size of the additional dimension. {\it Scale holography} is particularly useful for studying non-local theories, since the equivalent dual system on the higher dimensional manifold can be made to be local, as we illustrate with the specific example of the $p$-adic string theory.
\end{abstract}

\maketitle

\paragraph{Introduction.---} From the original concept of holography, proposed in 1993 by Gerard 't Hooft \cite{'tHooft:1993gx}, this idea has become ubiquitous in many different fields of modern theoretical physics. The pioneering discussions on black hole thermodynamics drove soon to the conjecture of the holographic `AdS/CFT' duality \cite{Maldacena:1997re}. These ideas have been intensively analyzed within a gravitational context. In short, they postulate that the study of a gravitating system defined on a given spacetime is equivalent to the study of a non-gravitating quantum system defined on a lower dimensional spacetime.
More precisely, the so-called holographic duality prescription establishes that vacuum correlators of a quantum field theory defined within a strong coupling regime in a $d$-dimensional spacetime can be computed by using appropriate fields in a $(d+1)$-dimensional anti-de Sitter spacetime \cite{Maldacena:1997re,Gubser:1998bc,Witten:1998qj}. Thermal states also can be studied by using the correspondence. In this case, the geometry of the $(d+1)$-dimensional spacetime is provided by an
anti-de Sitter black hole.

However, the diversity of fields with theoretical developments connected to holography is enormous and not restricted to the analysis of gravitational systems. The range of examples reaches from quantum chromodynamics \cite{Brodsky:2010ur, deTeramond:2008ht} to condensed matter \cite{Hartnoll:2008vx, McGreevy:2009xe,Cubrovic:2009ye}. Indeed, the importance of holography goes beyond particular realizations, and it is related to more foundational questions, which are not only interesting for a physics discussion but also from a philosophical approach \cite{Butterfield:2014oja, Butterfield:2010zz, Rickles:2013yka, Teh:2013xka}.
The duality between two systems   describing the same physics invites to discuss whether one of them can
be considered more fundamental or whether some elements of the theory are {\it just} emergent \cite{Seiberg:2006wf, Berenstein:2006yy,Domenech:2010nf,Verlinde:2010hp}.

In this work, we will present a new   scale-holographic relation between different systems.  More specifically, we will establish a holographic equivalence between a $d$-dimensional system and a whole family of systems described by Partial Differential Equations (PDE)  of different orders  in $(d+1)$ dimensions with appropriate initial and boundary conditions. The size of the additional dimension is related with the characteristic length scale of the original and hence the name.
The simplicity of the approach can be useful for discussing these questions. In addition, it provides a tool for solving involved PDEs such as those associated with  non-local systems as we will see with specific examples. 

Useful and insightful holographic duals can also be constructed for a variety of models that involve non-local kinetic operators. Such kind of kinetic operators appear for instance in superrenormalizable models for quantum gravity \cite{Modesto:2011kw,Modesto:2014lga}, in string field theory \cite{Calcagni:2013eua}, and in asymptotically safe theories of gravity \cite{Reuter:2001ag}, among many others.

We will use the $p$-adic string theory \cite{Freund:1987kt,Freund:1987ck,Frampton:1987sp,Brekke:1988dg} in a $(d+1)$-dimensional spacetime as a particularly interesting working example.
This model provides a tachyon effective action which reproduces correctly all the tree level bosonic string amplitudes on a D-brane and it is known to have a holographic correspondence with an effective string field theory in a $d$-dimensional spacetime \cite{Aref'eva:2014}.
Here we will show that this model is also holographically equivalent not only to a system described by the heat equation  in one additional dimension (this result is already well known and it has been thoroughly studied in Refs. \cite{Calcagni:2009jb,Calcagni:2007ef})
but also to a system associated with gravitational analogues. In both cases the appropriate choice of boundary conditions constitutes a fundamental piece of the correspondence as we will analyze in detail.

\paragraph{Scale-holographic duals.---} We will consider the following action for a real scalar field in $d$ spacetime dimensions  
\begin{equation}
S = -\int \diff^ dx\left[\dfrac{1}{2} \lambda_0^{-2}\phi\ F(\lambda_0^2\Box)\phi + V(\phi) \right],
\label{actionlambda}
\end{equation}
where $\Box$ is the $d$-dimensional D'Alembertian operator. We have added explicitly a constant with dimensions of length $\lambda_0>0$, so that $F$ is dimensionless, depends on the dimensionless operator $\lambda_0^2\Box$, and  is normalized $F(0)=1$. Then, $\phi$ has dimensions of   (length)$^{(2-d)/2}$.
For the case $F(0)=0$,  we can replace the kinetic function $F(z)$ with $H(z) = F(z) + 1$ 
and the potential $V(\phi)$ with $V_{\text{eff}}(\phi)= V(\phi) -\frac{1}{2} \lambda_0^{-2}\phi^2$. This new system satisfies $H(0)=1$. The case $F(0)=\infty$ will be dealt with later.
This action leads to the following equation of motion
\begin{equation}
F(\lambda_0^2\Box)\phi = -\lambda_0^2  V'(\phi).
\label{nonlocalEL}
\end{equation}

Let us introduce the new field $u(a,x)$, defined in $d+1$ dimensions as
\begin{equation}
u(a,x) = F(a\Box)\phi(x).
\label{udef}
\end{equation}
Since we have assumed that $F(0)=1$ we immediately see that $\phi$ is the initial condition in $a=0$ of the $(d+1)$-field $u$, i.e. $u(0,x)=\phi(x)$. 
Therefore, the differential equation of motion \eqref{nonlocalEL} acquires the algebraic form
\begin{equation}
u(\lambda_0^2,x) = -\lambda_0^{2}V'[u(0,x)],
\label{ubound}
\end{equation}
that relates the value of $u(a,x)$ at two different values of $a$, namely $a=\lambda_0^2$, the original length scale, and $a=0$. 

If we now differentiate Eq. \eqref{udef} with respect to $a$ we see that $u$ must satisfy the Partial Differential Equation (PDE)
\begin{equation}
\partial_a u(a,x) - \Box b(a\Box) u(a,x) = 0 ,
\label{ueq}
\end{equation}
where  the function $b(z)$ is just the logarithmic derivative of the kinetic function $F(z)$, that is,
$b = F'/F$.

To summarize, we have transformed the original problem into the following boundary problem
\begin{equation}
\left\lbrace \begin{array}{l}
\partial_a u(a,x) - \Box\ b(a\Box) u(a,x) = 0, \\[1ex]
u(\lambda_0^2,x) = -\lambda_0^2 V'[u(0,x)],
\end{array}\right.  
\label{equivaeq}
\end{equation}
where the first equation is a structural consequence of the definition of the $(d+1)$-field $u(a,x)$ and the second comes from the original equation of motion \eqref{nonlocalEL}. We will call this PDE boundary value problem  the \textit{scale-holographic dual} of the   original~\eqref{nonlocalEL}. 

We now prove that there is a one-to-one correspondence between the solutions of both problems. Indeed any solution of the original $d$-dimensional problem trivially leads via the definition \eqref{udef} to a solution of the $(d+1)$-dimensional scale-holographic dual. On the other hand, it is straightforward to see that the operator $F(a\Box)$ evolves in the extra dimension $a$ according to the differential operator equation $\partial_a F(a\Box) - \Box b(a\Box) F(a\Box)= 0$, with $F(0)=1$. Therefore $F(a\Box)$ generates the $a$-evolution according to Eq. \eqref{ueq}, i.e. any solution of  \eqref{ueq} must be of the form $F(a\Box)\phi(x)$ for some initial condition $\phi$. The boundary conditions of the scale-holographic dual problem ensure that $\phi$ is a solution of the original problem. Note also that, by construction, $(2\pi a)^{-d/2} \hat{G}\left( {x}/{\sqrt{a}} \right)$, where $\hat G(x)$ is the Fourier transform of $F(-k^2)$, is a fundamental solution of Eq.~\eqref{ueq}.

The scale-holographic dual of a non-linear system is still non-linear (the non-linearity being transferred to the boundary condition, which explicitly depends on the potential $V$) and no significant gain can be expected in this sense. However, for several kinetic functions,   the scale-holographic duals turn out to be much simpler.  Indeed, we will show below that the scale-holographic duals of several  non-local systems of physical relevance are local. More explicitly   a local theory defined on a certain volume with boundary conditions, when restricted only to (part of) its boundary, can reduce to a non-local theory, thus establishing the above-defined  duality between a non-local system in $d$ dimensions, and a local one in $d+1$ dimensions.
In the opposite direction, we can view this duality as providing a scale for the lower dimensional system
associated with the size of the extra dimension. This mechanism is analogous to the standard compactification used to obtain four observable dimensions in string theories \cite{cuerdas}. 
This method can also be used for kinetic functions not defined in 0. When doing radiative corrections, it is common to obtain kinetic functions that behave in certain energy scales as $\Box^{-1}$ or $\Box^{-2}$ \cite{codello1,codello2,cembra}. The problem with these kinetic functions is that we can no longer apply the method developed here, since $F(0)=\pm\infty$. However, this kinetic functions usually vanish at infinity, or at least tend to a constant value that can be normalized. Then, we can recover the original function, either with $F$ or $H$, as previously defined, not for $a=0$ but for $a\rightarrow \infty$.
Then, for these cases 
the holographic dual is given by \eqref{equivaeq} with the right hand side of the second equation evaluated at $a=\infty$ and not at $a=0$.

Let us now analyze some specific cases of physical interest.

\paragraph{The free particle.---}
Let us start with the simple case of a massive free particle described by the action
\begin{equation}
S=-\int \diff^dx\ \dfrac{1}{2}\ \phi (-\Box+m^2)\phi,
\end{equation}
for which $F(z) = -z+1$ and $\lambda_0^{-2}=m^2$. Then  $u$ satisfies the following PDE boundary value problem in $d+1$ dimensions:
\begin{equation}
\left\lbrace \begin{array}{l}
(1-a\Box)\partial_a u(a,x) + \Box u(a,x) = 0,\\[1ex]
u(\lambda_0^2,x) = 0 ,
\end{array} \right.  
\end{equation}
Note that we have applied $(1-a\Box)$ on the resulting equation in order to eliminate non-local operators. Although this operation could lead to additional spurious solutions, this is not the case as we will see, since the solutions do not lie in the kernel of $(1-a\Box)$.
We can solve this equation in momentum space. The solution has the following form
\begin{equation}
\hat{u}(a,k) = B(k)(1+ak^2),
\end{equation}
with $B(k)$ being determined by the boundary condition, which implies that 
\begin{equation} B(k)= \delta\left(1+\lambda_0^{2}k^2 \right)A(k) ,
\end{equation}
where $A(k)$ is an arbitrary function. Transforming back to position space and evaluating at $a=0$ we obtain   the solution to the original problem,
\begin{equation}
\phi(x) = u(0,x) = \dfrac{1}{(2\pi)^{d/2}}\int  \diff^dk  A(k)  \delta\left(1+\lambda_0^{2}k^2 \right) e^{ikx},
\end{equation}
which is the standard ---explicitly Lorentz invariant---solution in terms of plane waves, as expected.

For the massless case, $F(z)=-z$. As discussed above, the process is entirely analogous, now with $H(z)=F(z)+1$ and $V_{\text{eff}} = -\frac{1}{2}\lambda_0^{-2} \phi^2$. The solution is 
$\hat{u}(a,k) = A(k)\delta(k^2)$, 
with $A(k)$ being an arbitrary function.

\paragraph{The $p$-adic string theory.---}
Let us consider now the non-trivial case of the $p$-adic string  defined (after an appropriate definition of fields and variables \cite{Biswas:2010yx, Biswas:2010xq, Biswas:2009nx}) by 
\begin{equation}
S = - \int \diff^dx \ \left[ \dfrac{1}{2} \phi\ e^{\lambda_0^2 \Box}\phi -  \dfrac{\Lambda}{p+1}\phi^{p+1}\right].
\label{eq:padicaction}
\end{equation}
Here $p>1$ and $\Lambda$ has dimensions of (length)$^{(1-p)(d-2)/2}$.
Then the kinetic function is $F(z) = e^{z}$, and the equation of motion is
\begin{equation}
e^{\lambda_0^2 \Box}\phi(x) = \Lambda\phi^p(x),
\end{equation}
a PDE which is both non-linear and non-local.
Its   scale-holographic dual problem turns out to be a second-order PDE with a non-linear boundary condition as can be seen straightforwardly:
\begin{equation}
\left\lbrace \begin{array}{l}
\partial_a u - \Box u = 0, \\
u(\lambda_0^2,x) = \Lambda u^p(0,x), \qquad  \phi(x) = u(0,x).
\end{array}\right. 
\label{eq:padicdual}\end{equation}
We can perform now a Wick rotation on the time variable $x^\mu = (t,\bm x) \rightarrow x_\textsc{e}^a=(i\tau,\bm x)$, so that  $\Box\to\Box_\textsc{e}=\partial_\tau^2 + \nabla^2$ and transform this equation into a $(d+1)$-dimensional heat equation.
It should be noted that this relation between a $p$-adic string and the heat equation has already been made by some authors (see e.g. \cite{nardelli1,nardelli2,vladimirov2}) but it has not been solved yet without an explicit \textit{Ansatz}, not even in one dimension.

We will now provide a general method to find solutions of the (Wick-rotated) holographic dual \eqref{eq:padicdual} in $1+1$ dimensions, i.e. for the $p$-adic string action \eqref{eq:padicaction} in one dimension. This dual has the form
\begin{equation}
\left\lbrace 
\begin{array}{l}u_a-u_{\tau\tau}= 0 \\
u(\lambda_0^2,\tau) = \Lambda u^p(0,\tau) \qquad\end{array}\right. 
\end{equation}
An analysis of the Lie symmetry group structure of the heat equation  \cite{olver} shows that, given a solution $u_0(a,\tau)$, we can obtain a 6-parameter family of solutions acting on it with the symmetry group \cite{olver}:
\begin{align}
u &= u_1(a,\tau)\times u_0\left(\dfrac{e^{-\epsilon_4}(\tau-\epsilon_3 a)}{1-\epsilon_1 \tau}-\epsilon_5, \dfrac{e^{-2\epsilon_4}a }{1-\epsilon_1 a/4}-\epsilon_6 \right)
\nonumber\\
u_1 &= \dfrac{e^{\epsilon_2}}{\sqrt{1-\epsilon_1 a}}\text{exp}\left(  - \dfrac{\epsilon_3 \tau -\epsilon_1 \tau^2/4 - \epsilon_3^2 a}{1-\epsilon_1 a} \right).
\label{eq:sol-p-adic-heat}
\end{align}
If the seed solution $u_0$ is constant only three of these parameters are relevant. More explicitly, given the trivial solution $u_0=1$, the parameters $\epsilon_4,\epsilon_5,\epsilon_6$ are irrelevant and the requirement that $u$ satisfies the boundary condition implies that only one parameter $\tau_0$ is arbitrary:
\begin{equation}
\epsilon_1 = \dfrac{p-1}{p\lambda_0^2}, 
\quad 
\epsilon_2 = \dfrac{\ln p-2\ln \Lambda}{2(p-1)} +\dfrac{1}{4} \epsilon_1 \tau_0^2,
\quad 
\epsilon_3=\dfrac{1}{2}\tau_0 \epsilon_1.
\end{equation}
Finally, the solution to the original $p$-adic problem can be obtained Wick-rotating back this solution and evaluating it at $a=0$:
\begin{equation}
\phi(t) =u_1(0,t) = \left(\dfrac{\sqrt{p}}{\Lambda} \right)^{\frac{1}{p-1}} \exp\left[- p\,\epsilon_1 (t-t_0)^2 \right],
\end{equation}
where $t_0\equiv i \tau_0$,
which had already been obtained in  \cite{vladimirov2,nardelli1}.

\paragraph{Higher-order scale-holographic duals.---}
So far, we have mapped our very general system into a PDE boundary problem that is of first order in an emergent dimension, whose range is related to the scale of the original problem. However, we could have mapped our original system to a higher order one too.
Let us consider for instance the $p$-adic string theory again. As we said before, the kinetic function satisfies $F'(z)-F(z)=0$. However, this kinetic function satisfies a whole family of ODE of the form 
\begin{equation}
F^{(q)}(z)-F(z) = 0.
\end{equation}
The question now is whether or not we can map our original non-local system into more than one PDE boundary problem. To illustrate the answer, let us consider the case $q=2$. 
As we discussed before, $F(a\Box)$ is the $a$-evolution operator of the scale-holographic dual PDE of our problem. However, for a second order differential equation, there is not one propagator but two. The general solution of our scale-holographic dual will be a linear combination of the initial condition evolved with both propagators. To obtain at the end just one solution, another condition must be imposed to the problem.
Now, to obtain only the solution we demand $F$ to satisfy
\begin{equation}
\left\lbrace \begin{array}{l}
F''(z) - F(z) = 0,\\
F(0) = 1, \quad F'(0)=1 .
\end{array} \right. 
\end{equation}
Then, the only possible solution for the problem will be $F(z)  = e^z$, which is the kinetic function we are looking for. By defining again $u$ as $u(a,x) = F(a\Box)\phi(x)$, we obtain the following PDE boundary problem for $u$:
\begin{equation}
\left\lbrace 
\begin{array}{l}
\partial_a^2 u(a,x) -\Box^2 u(a,x) = 0, \\
u(\lambda_0^2,x) = \Lambda u^p(0,x), \quad \left. \partial_a u(a,x) \right\rvert_{a=0} = \Box u(0,x).
\end{array}
\right.  
\label{pde2}
\end{equation}
We see, that with the method developed in this work, we can map the original problem into a whole family of higher dimensional problems.
In the specific case of the non-local $p$-adic string theory, it is not only equivalent to the heat equation with specific boundary conditions but it is also equivalent (in two spatial dimensions and after performing a Wick's rotation) to gravitational analogues in condensed matter systems described by Eq. \eqref{pde2} \cite{barcelo}.
In this last case, not only boundary conditions but also initial conditions in the additional dimension need to be fixed as required by the scale holographic duality \eqref{pde2}.

In the case of a general kinetic operator $F(a\Box)$, for each order, there is a one-to-one equivalence between the original system and its scale holographic dual. Indeed the prove that we gave for the lowest order can be generalized straightforwardly to higher orders.

\paragraph{Conclusions.---} By extracting explicitly the length scale of the system, we have provided a method for mapping a very general system into a family of systems defined with an additional emergent dimension. This new holographic family allow us to interpret the solution of the original lower dimensional problem as the initial condition for the new higher dimensional system. These correspondence is particularly interesting for solving non-local problems since they can be solved
by studying their scale holographic duals. In this sense, it is interesting to realize that if the original non-local problem was non-linear, the resulting local higher dimensional boundary-value problem will also be non-linear. 

\acknowledgments 
We would like to thank M. A. Rodríguez, G. Álvarez, and G. Calcagni for fruitful conversations.
This work has been supported in part by the MINECO (Spain) projects 
FIS2014-54800-C2-2 (with FEDER contribution),
FIS2014-52837-P, FPA2014-53375-C2-1-P, and Consolider-Ingenio MULTIDARK CSD2009-00064.
J.A.R.C. acknowledges financial support from the {\it Jose Castillejo award} (2015).

\bibliography{biblio}

\end{document}